\documentclass[aps,prb,10pt,twocolumn,superscriptaddress,nobibnotes]{revtex4-1}

\usepackage{lmodern}
\usepackage[T1]{fontenc}
\usepackage[utf8]{inputenc}
\usepackage{graphicx}
\usepackage{amsmath,amssymb,amsfonts}
\usepackage[unicode=true,
            colorlinks=true,
            linkcolor=blue,
            citecolor=blue,
            urlcolor=blue]{hyperref} 
\usepackage{siunitx}
\usepackage{multirow}
\usepackage{lipsum}
\usepackage[x11names]{xcolor}

\newcommand{\domark}{%
  \vbox to 0pt{
    \kern-\dp\strutbox
    \smash{\llap{\color{red!90!black}\#\kern0.5em}}
    \vss
  }%
}

\begin{document}

\title{Near-unity light absorption in a monolayer $\mathrm{WS}_2$ van der Waals heterostructure cavity
}

\author{Itai Epstein}
\thanks{These authors contributed equally}
\affiliation{ICFO-Institut de Ciencies Fotoniques, The Barcelona Institute of Science and Technology, 08860 Castelldefels (Barcelona), Spain}
\email{itai.epstein@icfo.eu}

\author{Bernat Terr\'{e}s}
\thanks{These authors contributed equally}
\affiliation{ICFO-Institut de Ciencies Fotoniques, The Barcelona Institute of Science and Technology, 08860 Castelldefels (Barcelona), Spain}

\author{Andr\'{e} J. Chaves}
\affiliation{Grupo de Materiais Semicondutores e Nanotecnologia and Departamento de F\'{i}sica, Instituto Tecnol\'{o}gico de Aeron\'{a}utica, DCTA, 12228-900 S\~{a}o Jos\'{e} dos Campos, Brazil}

\author{Varun-Varma Pusapati}
\affiliation{ICFO-Institut de Ciencies Fotoniques, The Barcelona Institute of Science and Technology, 08860 Castelldefels (Barcelona), Spain} 

\author{Daniel A. Rhodes}
\affiliation{Department of Mechanical Engineering, Columbia University, New York, NY 10027} 

\author{Bettina Frank}
\affiliation{4th Physics Institute and Research Center SCoPE, University of Stuttgart, 70569 Stuttgart, Germany} 

\author{Valentin Zimmermann}
\affiliation{4th Physics Institute and Research Center SCoPE, University of Stuttgart, 70569 Stuttgart, Germany} 

\author{Ying Qin}
\affiliation{School for Engineering of Matter Transport and Energy, Arizona State University Tempe, AZ 85287, USA}

\author{Kenji Watanabe}
\affiliation{National Institute for Materials Science, 1-1 Namiki, Tsukuba 305-0044, Japan} 

\author{Takashi Taniguchi}
\affiliation{National Institute for Materials Science, 1-1 Namiki, Tsukuba 305-0044, Japan} 

\author{Harald Giessen}
\affiliation{4th Physics Institute and Research Center SCoPE, University of Stuttgart, 70569 Stuttgart, Germany}

\author{Sefaattin Tongay}
\affiliation{School for Engineering of Matter Transport and Energy Arizona State University Tempe, AZ 85287, USA}

\author{James C. Hone}
\affiliation{Department of Mechanical Engineering, Columbia University, New York, NY 10027}

\author{Nuno M. R. Peres}
\affiliation{Centro de F\'{i}sica and Departamento de F\'{i}sica and QuantaLab, Universidade do Minho, P-4710-057 Braga, Portugal}
\affiliation{International Iberian Nanotechnology Laboratory (INL), Av. Mestre José Veiga, 4715-330 Braga, Portugal}

\author{Frank H. L. Koppens}
\email{frank.koppens@icfo.eu}
\affiliation{ICFO-Institut de Ciencies Fotoniques, The Barcelona Institute of Science and Technology, 08860 Castelldefels (Barcelona), Spain}
\affiliation{ICREA – Instituci\'{o} Catalana de Recerca i Estudis Avan\c{c}ats, Barcelona, Spain}


\begin{abstract}
\textbf{
Excitons in monolayer transition-metal-dichalcogenides (TMDs) dominate their optical response and exhibit strong light-matter interactions with lifetime-limited emission. While various approaches have been applied to enhance light-exciton interactions in TMDs, the achieved strength have been far below unity, and a complete picture of its underlying physical mechanisms and fundamental limits has not been provided. Here, we introduce a TMD-based van der Waals heterostructure cavity that provides near-unity excitonic absorption, and emission of excitonic complexes that are observed at ultra-low excitation powers. Our results are in full agreement with a quantum theoretical framework introduced to describe the light-exciton-cavity interaction. We find that the subtle interplay between the radiative, non-radiative and dephasing decay rates plays a crucial role, and unveil a universal absorption law for excitons in 2D systems. This enhanced light-exciton interaction provides a platform for studying excitonic phase-transitions and quantum nonlinearities and enables new possibilities for 2D semiconductor-based optoelectronic devices.
}
\end{abstract}
\maketitle

The remarkable properties of excitons in monolayer TMDs, together with the ability to readily control their charge carrier density, have attracted a significant amount of interest in recent years. This has led to the observation of numerous phenomena\cite{Xu2014,Mueller2018}, such as higher order exciton complexes \cite{Mak2013,Ross2013,You2015,Plechinger2015}, coupled spin-valley physics \cite{Xiao2012,Zeng2012,Mak2012,Cao2012,Xu2014}, single photon quantum emitters \cite{Tonndorf2015,Chakraborty2015,Koperski2015,He2015,Srivastava2015}, together with monolayer semiconductor-based lasers, light-emitting-diodes, and photodetectors \cite{Radisavljevic2011,Yin2012,Lopez-Sanchez2013,Baugher2014,Ross2014,Ross2014}. Excitons in monolayer TMDs exhibit strong interaction with light, both in absorption and photoemission processes \cite{Mak2016,Wang2018,Mueller2018,Wurstbauer2017}, which facilitates large photovoltaic response enabled by strong peaks in the joint density of states \cite{Britnell2013} and strong-coupling \cite{Liu2015,Schneider2018}, to name two examples. Unlike their counterparts in quantum-well semiconductors, excitons in TMDs practically dominate the optical response of the material \cite{Wang2018,Mueller2018}. This stems from their large binding energies, which are a result of the strong Coulomb interaction and reduced screening that arise from their low dimensionality \cite{Chernikov2014}. The existence of robust excitonic states deep within the bandgap results in an optical bandgap that differs significantly from the electronic one \cite{Ugeda2014}, and thus prevails over the standard electronic-based optical response. \\

Nevertheless, in absolute values, the absorption of light by excitons in monolayer TMDs is far below unity, ranging between $2-12\%$ for as-transferred monolayers\cite{You2015,Poellmann2015,Robert2016,Okada2017,Li2014}, and about $20-30\%$ with the aid of a cavity \cite{Bahauddin2016,Liu2014a,Wang2017}. Similarly, $5-7\%$ absorption have been previously reported for trions \cite{Mak2013a,Zhang2014}. Using thicker TMDs, higher and broadband absorption has been observed, but the thickness was over $20$ monolayers \cite{Jariwala2016}. Thus, the question remains whether the achievable interaction strength can be pushed further and what would its limit be? Can unitary absorption be reached by excitons in an atomic thin layer? The answers may play an important role in the understanding of excitonic complexes in TMDs, and the realization of practical 2D material optoelectronic devices.      \\ 

Here, we demonstrate ultra-atrong light-exciton interaction strength in a $\mathrm{WS}_2$-based high quality van der Waals heterostructure cavity (VHC), i.e an optical cavity built from van der Waals materials, which can be controlled both electrically and optically. While the cavity is quite broadband, the near-unity absorption is attainable owing to four major elements: (1) Achievable narrow excitonic linewidths \cite{Cadiz2017,Ajayi2017}. (2) The ability to carefully balance the interplay between the radiative and non-radiative decay rates, $\gamma_{\mathrm{r}}$ and $\gamma_{\mathrm{nr}}$, respectively. (3) The enhancement of the vacuum radiative decay rate, $\gamma_{\mathrm{r,0}}$, via the Purcell factor, shifting the maximum absorption to larger (and attainable) linewidths.  (4) Obtaining extremely low dephasing rate, $\gamma_{\mathrm{d}} \ll \gamma_{\mathrm{r}}, \gamma_{\mathrm{nr}}$. \\

We show that this approach yields a large photo-excited excitonic population, with record values of $\sim92\%$ excitonic absorption, $\sim41\%$ for singlet and triplet trion states, and even the observation of the next negatively charged trion state with $\sim28\%$ absorption. In addition, it enables the observation of biexcitons photoluminescence (PL) at ultra-low continuous-wave (cw) laser powers down to a few $\mathrm{nW}$, which is three orders of magnitude lower than previously reported values \cite{Ye2018,Barbone2018}. We introduce an analytical approach to describe the light-exciton-cavity interaction, which is based on the semiconductor Bloch equations combined with a quantum transfer matrix method. The model takes into account the contribution of both the exciton radiative and non-radiative decay rates, $\gamma_{\mathrm{r}}$ and $\gamma_{\mathrm{nr}}$, which have already been shown to affect the excitonic coherence and high reflection of monolayer TMDs \cite{Scuri2018,Back2018,Zeytinoglu2017}. In addition, we include the existence of a pure dephasing rate, $\gamma_{\mathrm{d}}$, to account for quantum coherence effects as part of the multiple interferences within the cavity. We find that the relation between $\gamma_{\mathrm{r}}$ and $\gamma_{\mathrm{nr}}$ establishes the condition for maximal absorption, and its limit is set by the value of the pure dephasing rate $\gamma_{\mathrm{d}}$, basically limiting the coherence of the system. Experimentally, we control the non-radiative channels with temperature, and the radiative via the geometrical parameters of the VHC (Purcell effect). Finally, we demonstrate the existence of a universal absorption law for excitons in 2D systems in this class of devices. \\

\begin{figure*}[ht!] 
  \centering
  \includegraphics[scale=0.6]{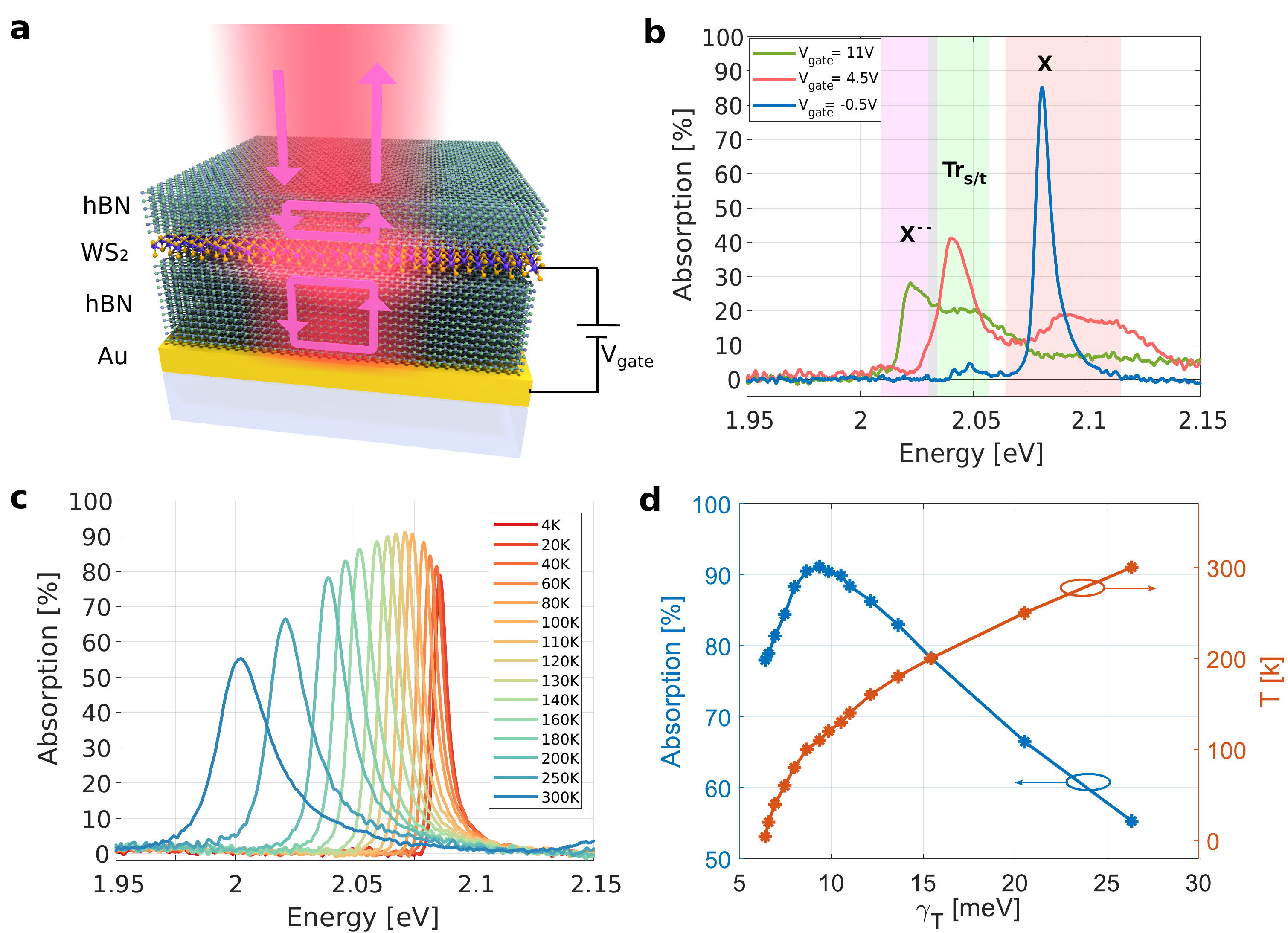} 
   \caption{
	\textbf{Light-exciton interaction in the van der Waals heterostructure cavity}. \textbf{a} The structure of the VHC - an hBN/monolayer $\mathrm{WS}_2$/hBN heterostructure with optimized hBN thicknesses is transferred on top of a gold back reflector. The gate voltage $\mathrm{V}_\mathrm{gate}$ controls the doping in the $\mathrm{WS}_2$.\textbf{b} Gate dependent spectral absorption for sample U2, showing $\sim85\%$ excitonic absorption, $\sim41\%$ trion absorption, and $\sim28\%$ of the next negatively charged trion state, at voltages $\mathrm{V}_\mathrm{gate}$=-0.5~V,  $\mathrm{V}_\mathrm{gate}$=4.5~V, and $\mathrm{V}_\mathrm{gate}$=11~V, respectively. \textbf{c} Temperature dependent excitonic spectral absorption from sample U1, showing the non-trivial behavior of the absorption, and a maximum absorption of $\sim92\%$ at $\mathrm{T}$=110~K. \textbf{d} The maximum excitonic absorption dependency on the total linewidth $\gamma_{\mathrm{T}}$ (blue curve) and the temperature dependent exciton  linewidth (red curve). }
	\label{fig:figure1}
\end{figure*}

The VHC is composed of a monolayer $\mathrm{WS}_2$ encapsulated by hexagonal Boron Nitride (hBN), and transferred on top of a gold back reflector (Fig.~\ref{fig:figure1}a). We have fabricated two such heterostructure cavities - sample U1 in which we used for the back reflector a single crystalline, atomically flat gold flake \cite{Podbiel2019}, and sample U2, where we used a standard evaporated gold film as the back reflector, which is also used for electrostatic gating. In these structures the optical transmission (for visible light) is zero, and the absorption can be obtained from $1-\frac{R}{R_0}$, where $R$ and $R_0$ are the reflection from the structure with and without the TMD, respectively. \\

Fig.~\ref{fig:figure1}b shows the typical absorption spectra, obtained from sample U2, for different gate voltages at $\mathrm{T}\mathrm{=4~K}$. An absorption value as high as $\thicksim85\%$ can be seen at an energy of $\mathrm{E}=2.08$ $\mathrm{eV}$, corresponding to the $\mathrm{WS}_2$ neutral exciton ($\mathrm{X}$), an absorption value of $\thicksim41\%$ at $\mathrm{E}=2.041$ $\mathrm{eV}$, corresponding to both singlet and triplet trions (${\mathrm{Tr}_{\mathrm{t}}/\mathrm{Tr}_{\mathrm{s}}}$) (see SI), and an absorption value of $\thicksim28\%$ at $\mathrm{E}=2.023$ $\mathrm{eV}$, corresponding to the next charged state of the trion ($\mathrm{X}^{--}$). All energetic positions of the absorption peaks and 	separations are in agreement with previous reports on $\mathrm{WS}_2$ \cite{Plechinger2016,Plechinger2015}. By changing the gate voltage we control the charge carriers in the $\mathrm{WS}_2$, and thus the relative spectral weights of ($\mathrm{X}$), ($\mathrm{T}_{\mathrm{r}}$) and ($\mathrm{X}^{--}$) \cite{Mak2013a, Paur2019}, i.e, the absorption spectrum can be controlled electrically. To the best of our knowledge, these are record absorption values together with the first observation of the $\mathrm{X}^{--}$ peak in an absorption spectra of TMDs. This is a direct result of the strong light-matter interaction provided by the VHC.\\

To study the ultimate absorption limits, we vary the temperature, which controls the excitonic linewidth, as presented in Fig.~\ref{fig:figure1}c,d. While the exciton linewidth shows a continuous decrease with decreasing temperature (Fig.~\ref{fig:figure1}d red curve), as known for the ground state exciton in semiconductors \cite{Rudin1990} and TMDs \cite{Cadiz2017,Selig2016}, the excitonic absorption shows a non-monotonic temperature dependence. An absorption value of $55\%$ can already be seen at room-temperature, which increases to a maximum value of $\sim92\%$ at $\mathrm{T}\mathrm{=110~K}$, and then decreases rapidly to $\sim77\%$ at $\mathrm{T}\mathrm{=4~K}$ (Fig.~\ref{fig:figure1}d blue curve), the low temperature limit of our cryostat. \\

In order to understand the physical origin of this behavior, we developed a theoretical formalism that combines an equation of motion method for the exciton, which is similar to the well-known (interacting) Bloch equations \cite{Combescot2009,Chaves2017}, together with a quantum treatment of the cavity using a quantum transfer matrix method (QTMM, see SI). To take into account the pure dephasing, the QTMM treats the electromagnetic fields as operators, rather than classical fields. This provides the relation between the polarization operator and the field in the 2D material. The solution, which describes the optical response of the TMD, leads to the an Elliott-type formula appropriate for the 2D material \cite{Chaves2017}. This allows us to calculate the absorption, via the reflection operator's expectation value, taking into account the contributions of both $\gamma_{\mathrm{r,0}}$ (and its Purcell enhancement), $\gamma_{\mathrm{nr}}$, and $\gamma_{\mathrm{d}}$. The latter is a quantum coherent effect that is important to consider due to the multiple interferences in the cavity, making the absorption sensitive to exciton dephasing \cite{Scuri2018}. \\

Following this approach (see SI), the maximum absorption can then be approximated by:

\begin{equation}
\mathrm{A}_\mathrm{max}=\xi_1 \frac{\gamma_{\mathrm{r,0}}}{\gamma_{\mathrm{T}}} \left[ 1-\xi_2(1+2\frac{\gamma_{\mathrm{d}}}{\gamma_{\mathrm{T}}})\frac{\gamma_{\mathrm{r,0}}}{\gamma_{\mathrm{T}}} \right], 
\label{eq:1}
\end{equation}\\
with 
\begin{equation}
\gamma_{\mathrm{T}}= \gamma_{\mathrm{nr}}+2\gamma_{\mathrm{d}}+\zeta \gamma_{\mathrm{r,0}}, 
\label{eq:2}
\end{equation}
where $\frac{\gamma_{\mathrm{T}}}{\gamma_{\mathrm{r,0}}}$ and $\frac{\gamma_{\mathrm{d}}}{\gamma_{\mathrm{T}}}$ are the normalized vacuum radiative decay and dephasing rates, and the coefficients $\xi_1$, $\xi_2$, and $\zeta$ are (see SI) parameters depending on the geometry, defined by the sizes and dielectric functions composing the different parts of the cavity, with $\zeta$ representing the Purcell factor, and $\gamma_{\mathrm{r}}=\zeta\gamma_{\mathrm{r,0}}$ being the renormalized radiative decay rate. \\

\begin{figure*}[ht!] 
   \centering
   \includegraphics[scale=0.6]{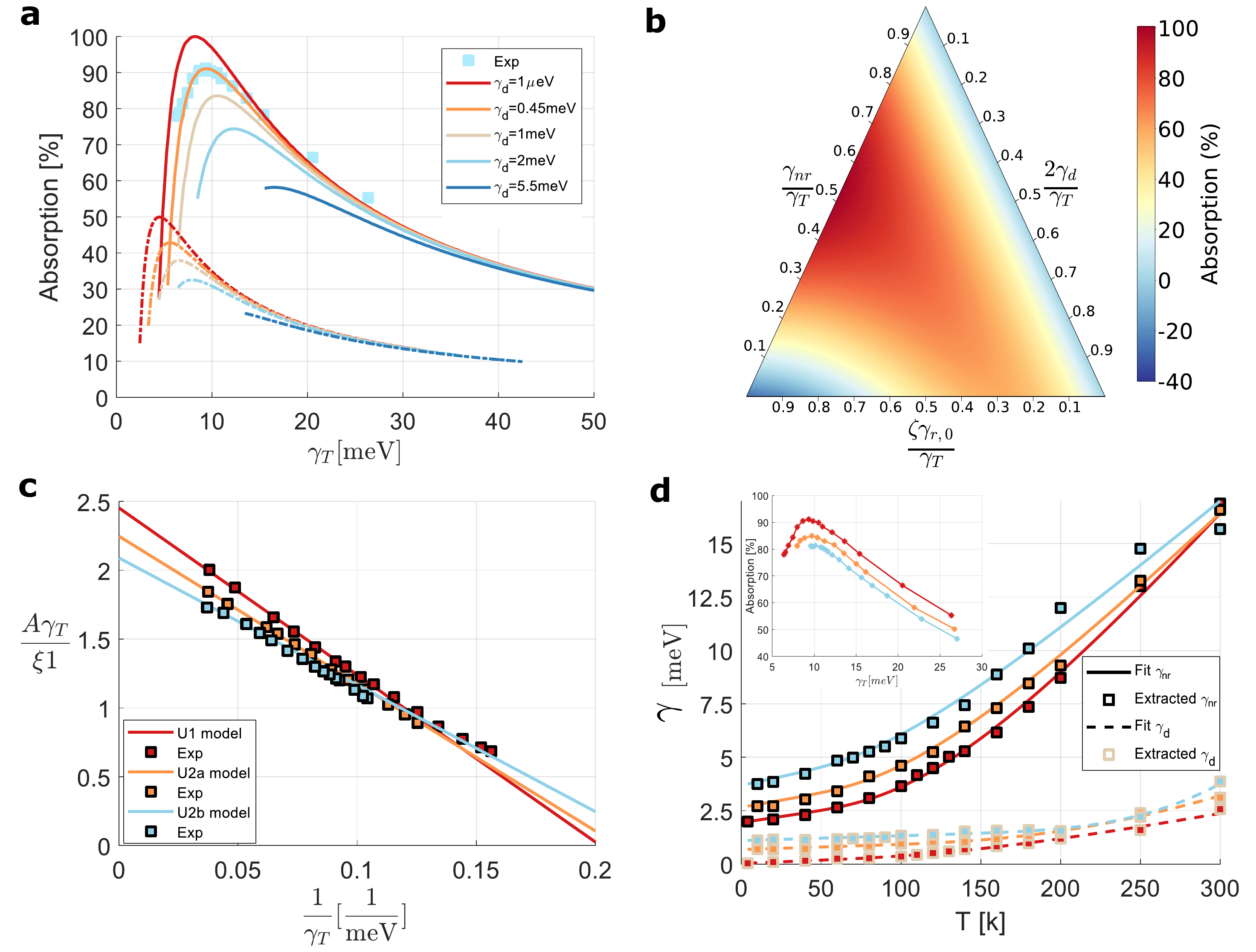} 
   \caption{
	\textbf{Analytical modeling of the light-exciton-cavity interaction.} \textbf{a} Calculated absorption dependence on $\gamma_{\mathrm{T}}$, for the VHC (solid lines) and a suspended TMD (dashed lines), for several dephasing values. The squares show the experimental results from sample U1. \textbf{b} Calculated maximal absorption as function of the three normalized radiative decay rates. $100\%$ absorption is achieved when $\gamma_{\mathrm{d}}=0$ and $\zeta\gamma_{\mathrm{r,0}}\approx\gamma_{\mathrm{nr}}$. The negative absorption observed at the bottom left corner is due to the dominance of $\gamma_{\mathrm{r}}$ and is associated with high reflection. \textbf{c} Temperature dependent experimental results for the two different samples, U1 and U2, fitted with Eq. \ref{eq:3} exhibiting the universal absorption law. The different radiative decay rates can also be seen at the crossing of the lines with the $y$-axis. U2a and U2b correspond to two different locations on sample U2 that were chosen due to different values of the total linewidth at 4K. \textbf{d} Extracted $\gamma_{\mathrm{nr}}$ and $\gamma_{\mathrm{d}}$ and their phenomenological fit, showing the correlation between the achievable absorption and the decay rates. 
	}
\label{fig:figure2}
\end{figure*}

To expose the roles of the different decay rates in the absorption behavior, we show in Fig.~\ref{fig:figure2}a the calculated absorption dependence on the exciton linewidth, $\gamma_{\mathrm{T}}$, for several dephasing values. For each $\gamma_{\mathrm{d}}$, $\gamma_{\mathrm{r,0}}$ is kept constant, as it does not depend on temperature, and we vary $\gamma_{\mathrm{nr}}$ (solid lines). For the simple case of negligible $\gamma_{\mathrm{d}}$, we obtain a matching condition for the maximum absorption point: $\gamma_{\mathrm{nr}}\approx\zeta\gamma_{\mathrm{r,0}}$ (see SI), and $100\%$ absorption. This implies that the relation between $\gamma_{\mathrm{r}}$, and $\gamma_{\mathrm{nr}}$, rather than their absolute values, sets the matching condition for maximal absorption, and $\gamma_{\mathrm{d}}$ sets the limit of the absolute achievable absorption when the matching condition is fulfilled. For comparison, the absorption of a suspended monolayer is presented for the same dephasing values (dashed curves), exhibiting the same behavior and showing the known maximal absorption limit of $50\%$ for a thin layer\cite{Thongrattanasiri2012}, but at smaller and less attainable $\gamma_{\mathrm{T}}$. Thus, the VHC not only enables near-unity absorption, but via the Purcell effect, also shifts the matching condition to larger $\gamma_{\mathrm{T}}$, making the experimental realization of such large absorption easier to implement. \\

For the more general case including finite dephasing, the matching condition translates to a more complex relation between $\frac{\gamma_{\mathrm{r,0}}}{\gamma_{\mathrm{T}}}$ and $\frac{\gamma_{\mathrm{d}}}{\gamma_{\mathrm{T}}}$ (Fig.~\ref{fig:figure2}b). A ternary plot of the absorption as function of the normalized radiative, nonradiative and pure dephasing decay rates, is presented in Fig.~\ref{fig:figure2}b, showing all the possibilities in the decay rates' space. It can be seen from Fig.~\ref{fig:figure2}a that away from the matching condition, the dephasing has little effect and the absorption decreases following two different regimes. For very small $\gamma_{\mathrm{T}}$, $\gamma_{\mathrm{r,0}}$ dominates the linewidth and the absorption decreases as the TMD becomes highly reflective (corresponding to negative absorption, as seen in (Fig.~\ref{fig:figure2}b)) \cite{Scuri2018,Back2018,Zeytinoglu2017}. For very large $\gamma_{\mathrm{T}}$, $\gamma_{\mathrm{nr}}$ and $\gamma_{\mathrm{d}}$ dominates the linewidth and the absorption is decreased as the TMD becomes transparent.\\

The above discussed decay rates are highly dependent on the cavity design, TMD quality, fabrication-induced interface quality, and can vary locally\cite{Back2018}. Yet, via a simple representation of Eq. \ref{eq:1} as function of $\gamma_{\mathrm{T}}^{-1}$, the model makes a striking prediction and inescapable universal feature of this class of devices: 
\begin{equation}
\frac{\mathrm{A}_{\rm{max}}\gamma_{\mathrm{T}}}{\xi_1}= \gamma_{\mathrm{r,0}}\left[1-\xi_2\left(1+2\frac{\gamma_{\mathrm{d}}}{\gamma_{\mathrm{T}}}\right) \frac{\gamma_{\mathrm{r,0}}}{\gamma_{\mathrm{T}}} \right]. 
\label{eq:3}
\end{equation}
This implies a linear relationship between $\mathrm{A}_{\mathrm{max}}\gamma_{\mathrm{T}}/\xi_1$ and $1/\gamma_{\mathrm{T}}$, provided that $\gamma_{\mathrm{d}}/\gamma_{\mathrm{T}}\ll 1$. Indeed, the universal law is confirmed by the experimental data from different samples and locations presented in Fig.~\ref{fig:figure2}c. The different samples follow their own straight line, which encodes their different quality, but the generic behavior is the same for all. In addition, $\gamma_{\mathrm{r,0}}$ of the different samples can be extracted from the (extrapolated) crossing point $1/\gamma_{\mathrm{T}}=0$. In principle, this universal law should hold for any 2D excitonic system where $\gamma_{\mathrm{d}} \ll \gamma_{\mathrm{T}}$. \\

Another important outcome of this analysis is the ability to compare different excitonic properties via their absorption response. Fig.~\ref{fig:figure2}d shows the extracted temperature dependent $\gamma_{\mathrm{nr}}$ and $\gamma_{\mathrm{d}}$ (see SI) and their phenomenological fit \cite{Selig2016,Scuri2018}. These correlate directly with their absorption behavior, indicating that higher dephasing leads to lower absorption. \\

The light passing through the heterostructure with a back reflector can be regarded as an asymmetric Fabry-Perot cavity. The simplest case of such a cavity is a single dielectric layer on top of a mirror, known as the Salisbury screen, and when the thickness of the dielectric is chosen to be a quarter wavelength of the light in the dielectric, constructive interference takes place on the surface of the dielectric and the field is locally enhanced. Specifically for the VHC, the situation is more complex owing to the contributions from the multiple-layer structure and the gold mirror. Due to the penetration depth into the gold, the bottom hBN thickness is less than the quarter wavelength, and the top hBN is then optimized. In addition, the TMD, with its tunable properties, acts as a layer within the cavity itself, which can be either absorptive or reflective. On the other hand, the cavity provides a degree of freedom to balance the matching condition in order to achieve strong light-exciton interaction. \\

\begin{figure*}[ht!] 
   \centering
   \includegraphics[scale=0.6]{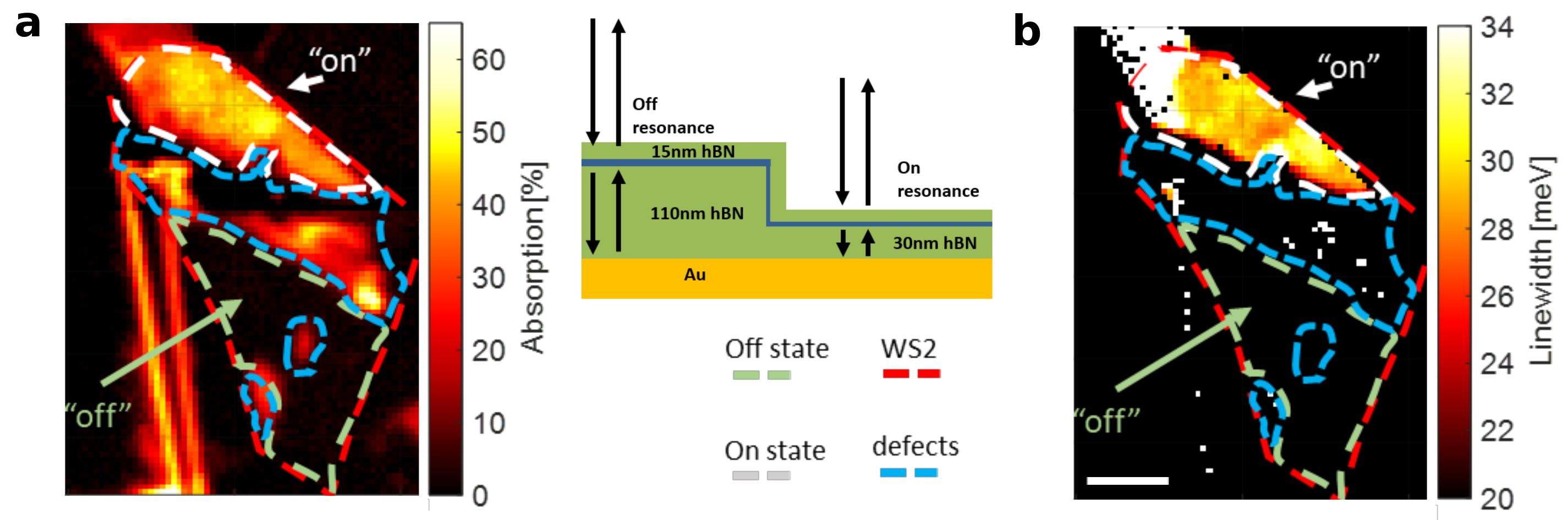} 
   \caption{
	\textbf{Tuning the excitonic optical response via the geometrical design of the VHC}. \textbf{a} Spatial distribution of the absorption at room temperature showing the ''$\mathrm{on}$''-state for the area of maximal interaction and ''$\mathrm{off}$''-state at the area of minimal interaction. The schematic shows the structure of the sample with two different bottom hBN thicknesses. \textbf{b} Spatial distribution of the measured linewidth of the excitonic absorption in \textbf{a}, showing a clear correlation between linewidth and absorption. The scale bar represents $5~\mu \mathrm{m}$.  
	}
\label{fig:figure3}

\end{figure*}

In the same manner we have designed the VHC for maximal interaction strength, it can be designed to any intermediate value, and even to completely turn off the interaction between the light and the TMD. To demonstrate this, we fabricated on the same device two different cavities, one for which the interaction is optimized (denoted as ''$\mathrm{on}$''), and one for which the interaction is minimized (denoted as ''$\mathrm{off}$''). This is done by adding another hBN flake below a part of the heterostructure. The spatial distribution of the absorption for this device at $\mathrm{T} \mathrm{=300~K}$ is presented in Fig.~\ref{fig:figure3}a and its extracted exciton linewidth distribution in Fig.~\ref{fig:figure3}b. The spatial correlation between the absorption and linewidth can be directly observed in the ''$\mathrm{on}$'' areas of the two figures, i.e., lower linewidth correlates to higher absorption, which is indeed the case for $\mathrm{T}\mathrm{=300~K}$ (see Fig.~\ref{fig:figure1}d). The source of the spatial distribution comes from the inhomogeneity of the sample, a well known issue in TMDs \cite{Back2018}. Furthermore, the extracted $\gamma_{\mathrm{r}}$ from the two cavities yields $2.2$ $\mathrm{meV}$ and $70$ $\mu\mathrm{eV}$ in the ''$\mathrm{on}$'' and ''$\mathrm{off}$'' regions, respectively. This is in agreement with the Purcell effect linewidth modulation obtained by Ref \cite{Fang2019}.  \\

The ability to achieve near-unity excitonic absorption implies that a large photo-excited excitonic density can be obtained, while maintaining low excitation power. It was already shown that TMD's are highly affected by the excitation power, resulting in either heating effects that changes the excitonic properties temporarily \cite{Currie2015}, or permanent effects that completely alter the material's response, such as optical doping and environmental surface interactions \cite{You2015,Cadiz2016}. Yet, this challenging and desirable high exciton density plays a major role in several physical phenomena, such as Bosonic condensation, phase transitions\cite{Arp2019} and biexcitons emission\cite{You2015}, for example. The formation of biexcitons is directly related to high excitonic density and thus also to the possible appearance of the biexciton peak in the PL spectrum \cite{You2015}. The predicted intensity relation between the exciton and biexciton emission is a power law - $I_{\mathrm{XX}}=I_{\mathrm{X}}^{\alpha}$, with $\alpha$ ranging between $1.2-1.9$ due to lack of thermal equilibrium \cite{You2015}. Thus, the latter can be used to probe the efficiency of exciton photo-generation in the VHC. \\

\begin{figure*}[ht!] 
   \centering
   \includegraphics[scale=0.57]{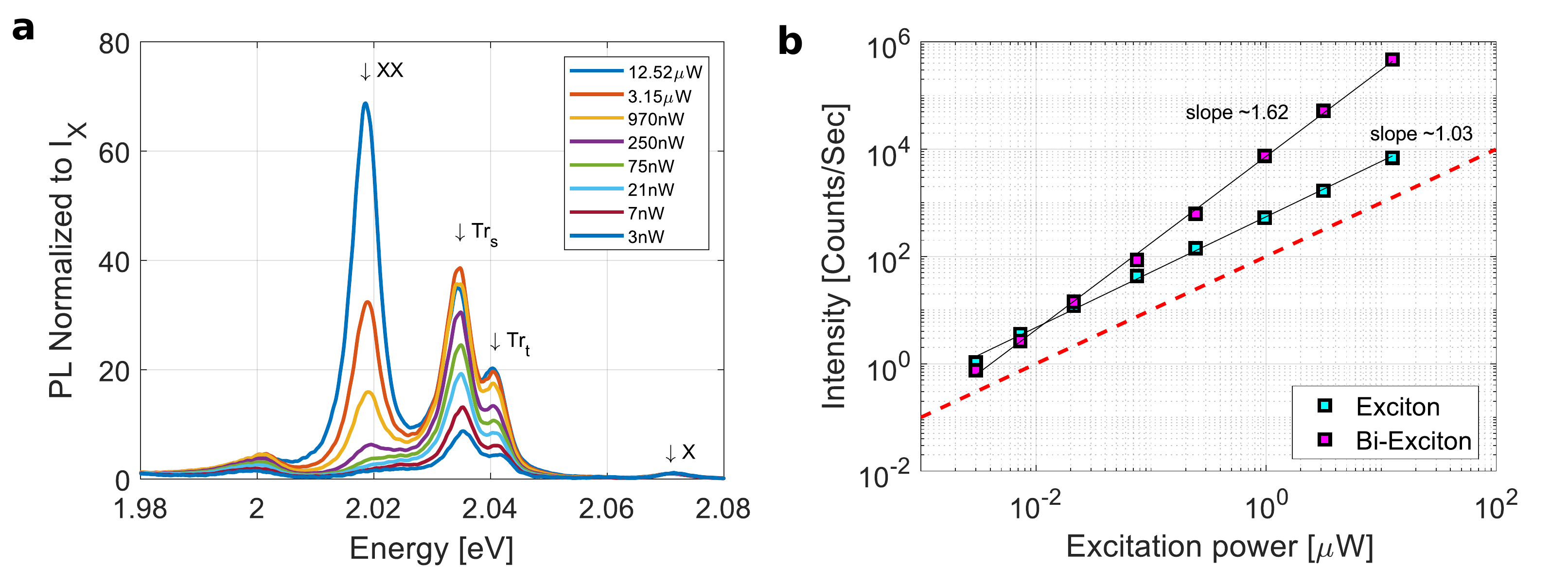} 
   \caption{
	\textbf{Highly efficient biexciton emission}. \textbf{a} Normalized PL spectra for different cw excitation power ranging from $\mu \mathrm{W}$ to $\mathrm{nW}$. \textbf{b} Linear fits of the exciton and biexciton PL emission showing the superlinear behavior of the biexcitons peak. The dashed red line correspond to $\mathrm{slope}=1$.
	}
\label{fig:figure4}
\end{figure*}

Fig.~\ref{fig:figure4}a shows the PL spectra, normalized to the exciton emission intensity ($I_{\mathrm{X}}$), obtained from the VHC for different cw excitation powers. Several peaks can be observed and are marked as $\mathrm{X}$ - exciton at $\mathrm{E}=2.071$ $\mathrm{eV}$, $\mathrm{Tr}_{\mathrm{t}}/\mathrm{Tr}_{\mathrm{s}}$ - singlet/triplet trions \cite{Plechinger2016} at $2.034$ $\mathrm{eV}/2.04$ $\mathrm{eV}$, respectively, and $\mathrm{XX}$ - biexciton at $\mathrm{E}=2.018$ $\mathrm{eV}$. Remarkably, the $\mathrm{XX}$ emission peak can be observed at excitations powers down to few $\mathrm{nW}$. This excitation power is three orders of magnitude smaller than the lowest previously reported for biexcitons \cite{Ye2018}. Actually, for excitation powers above $30$ $\mu \mathrm{W}$ the $\mathrm{XX}$ emission intensity is so high that it saturates our detector. Fig.~\ref{fig:figure4}b shows the power law analysis of the $I_{\mathrm{XX}}$ and $I_{\mathrm{X}}$ peaks, and the obtained $\alpha=1.62$ for $I_{\mathrm{XX}}$ confirms the identity of the biexciton. From the maximal used power of $400$ $\mu \mathrm{W}$ we can calculate an excitonic density of $\thicksim2.5\cdot10^{11}~\mathrm{cm}^{-2}$. This exciton density is 2 orders of magnitude larger than what was achieved in previous works using cw excitation, and is similar to densities obtained using ultrafast laser pulses \cite{Poellmann2015}. While it is possible that at high cw excitation powers heating effects may kick in, using pulsed excitation (as in ref \cite{Poellmann2015} for example) with our VHC, one can obtain an excitonic density of $\thicksim5\cdot10^{13}~\mathrm{cm}^{-2}$ with little heating, as compared to $\thicksim6.4\cdot10^{12}~\mathrm{cm}^{-2}$. \\ 

In conclusion, the new type of optical cavity presented here, which is based solely on van der Waals heterostructures, is designed for ultra-strong light-exciton interaction, and unveils a universal absorption law for excitons in 2D systems. This enhanced light-exciton interaction can act as a platform for probing quantum nonlinear dynamics of excitons \cite{Zeytinoglu2017, Scuri2018} and their state of matter \cite{Arp2019}, and paves the way to efficient optoelectronic devices, such as detectors, modulators and optically-pumped light emitting devices, based on monolayer semiconductors.\\

\section*{\textbf{Methods}}
\textbf{Optical setup}
All temperature measurements were done in an Attodry800 cryostat, and spectral detection with an Andor spectrometer. A white light source was used for the reflection measurements and a $532$ $\mathrm{nm}$ laser for the PL. The light was focused using a Nikon objective with $\mathrm{NA}=0.6$.

\textbf{Synthesis}
(sample U2)\indent WS$_2$ was grown using chemical vapor transport with a WCl$_6$ precursor in excess sulfur. W, 99.999 \%, and sulfur (99.9995 \%) were first loaded into a quartz ampoule in a 1:2, W:S, ratio with an additional excess of 24 mg/cm$^3$ of sulfur. 60 mg of WCl$_6$ was added into the quartz ampoule inside of a nitrogen glovebox, the ampoule was then removed from the glovebox and quickly pumped down to 5 x 10$^{-5}$ Torr to avoid deterioration of the WCl$_6$. The ampoule was then sealed under vacuum.  The reagents were subsequently heated to 1000 $^{\circ}$C within 24 h and with a 100 $^{\circ}$C temperature gradient between the hot and cold zone. The ampoule was then allowed to dwell at this temperature for 2 weeks before being cooled to 400 $^{\circ}$C over an additional 2 weeks. The as harvested crystals were then rinsed in acetone and isopropanol and dried in air before being used.   

\textbf{Device Fabrication}
Monolayer flakes of WS2 were mechanically exfoliated from the aforementioned single crystals on to 285 nm SiO$_2$ chip. On a separate chip, hexagonal boron nitride (hBN) was exfoliated and picked up using a dry transfer, PDMS/polypropylene carbonate (PPC) stamp method mounted on to a glass slide. Subsequently the hBN is used to pick up, in order, two few-layer graphene contacts, the monolayer WS$_2$, and an underlying hBN. The final stack is then transferred on to a the gold pad at 125 $^{\circ}$C and rinsed in acetone overnight.

\section*{\textbf{Author Contribution}}
I.E. conceived the idea, performed simulations, experiments and analysis of the results. B.T. fabricated samples and analyzed the results. V.P. assisted in measurements. A.C. and N.P introduced the theoretical framework and simulations. D.R., B.F., V.Z., Y.Q, K.W., T.T., H.G., S.T., and J.H. supplied materials and/or samples. F.H.L.K. supervised the work. All authors contributed to discussions and writing of the manuscript. 

\begin{acknowledgments}
{ \small
I.E. thanks Mr. David Alcaraz Iranzo, Dr. Fabien Vialla, and Dr. Antoine Reserbat-Plantey for fruitful discussions. F.H.L.K. acknowledges financial support from the Government of Catalonia trough the SGR grant (2014-SGR-1535), and from the Spanish Ministry of Economy and Competitiveness, through the “Severo Ochoa” Programme for Centres of Excellence in R and D (SEV-2015-0522), support by Fundacio Cellex Barcelona, Generalitat de Catalunya through the CERCA program,  and the Mineco grants Ramon y Cajal (RYC-2012-12281,  Plan Nacional (FIS2013-47161-P and FIS2014-59639-JIN) and the Agency for Management of University and Research Grants (AGAUR) 2017 SGR 1656.  Furthermore, the research leading to these results has received funding from the European Union Seventh Framework Programme under grant agreement number 785219 Graphene Flagship. This work was supported by the ERC TOPONANOP under grant agreement number 726001 and the MINECO Plan Nacional Grant 2D-NANOTOP under reference number FIS2016-81044-P. S.T acknowledges support from NSF DMR-1552220 and DMR-1838443. NMRP acknowledges financing from European Commission through the project ”Graphene-Driven Revolutions in ICT and Beyond” (Ref. No. 785219)and from FEDER and the Portuguese Foundation for Science and Technology (FCT) through project POCI-01-0145- FEDER-028114. H.G. and B.F. acknowledge support from ERC advanced grant COMPLEXPLAS.
} \end{acknowledgments}

\bibliography{Unity_abs}

\end{document}